\documentclass[prl,twocolumn,amsmath,amssymb,amsfonts,footinbib]{revtex4-2}

\usepackage{epsfig}
\usepackage{graphicx}
\usepackage{grffile} 
\usepackage{dcolumn}
\usepackage{bm}
\usepackage[titletoc]{appendix}

\usepackage{graphicx}
\usepackage{subfigure}
\usepackage{amsfonts}
\usepackage{amsgen}
\usepackage{amsbsy}
\usepackage{amsmath}
\usepackage{amssymb}
\usepackage{mathrsfs}
\usepackage{mathtools}
\usepackage{epstopdf}
\usepackage{caption}
\usepackage{tabularx}
\usepackage{tabu}
\usepackage{xcolor}
\usepackage{nicefrac}
\usepackage{enumitem}
\usepackage{gensymb}
\usepackage{natbib}
\usepackage{upgreek}
\usepackage{textcomp}  

\newcommand\etal{et al.~}

\newcommand\ie{\textit{i.e.~}}

\newcommand\eg{e.g.~}

\begin{document}

\title{Irreversibility and rate dependence in sheared adhesive suspensions}

\author{Zhouyang Ge$^{1,2}$}\email{zhoge@mech.kth.se}
\author{Raffaella Martone$^3$}
\author{Luca Brandt$^1$}
\author{Mario Minale$^3$}\email{mario.minale@unicampania.it}

\affiliation{$^1$ Department of Engineering Mechanics, KTH Royal Institute of Technology, SE-100 44 Stockholm, Sweden}
\affiliation{$^2$ Department of Mechanical Engineering, University of British Columbia, Vancouver, BC, V6T 1Z4, Canada}
\affiliation{$^3$ Department of Engineering, University of Campania ``Luigi Vanvitelli'', Real Casa dell'Annunziata, via Roma 29-81031 Aversa, CE, Italy}

\date{\today}

\begin{abstract}

Recent experiments report that slowly-sheared noncolloidal particle suspensions can exhibit 
unexpected rate($\omega$)-dependent complex viscosities in oscillatory shear, despite a constant relative viscosity in steady shear.
Using a minimal hydrodynamic model, we show that a weak interparticle attraction reproduces this behavior.
At volume fractions $\phi=20\sim50$\%, the complex viscosities in both experiments and simulations display power-law reductions in shear,
with a $\phi$-dependent exponent maximum at $\phi=40$\%, resulting from the interplay between hydrodynamic, collision and adhesive interactions.
Furthermore, this rate dependence is accompanied by diverging particle diffusivities and pronounced cluster formations
even at small oscillation amplitudes $\gamma_0$. 
Previous studies established that suspensions transition from reversible absorbing states to irreversible diffusing states 
when $\gamma_0$ exceeds a $\phi$-dependent critical value $\gamma_{0,\phi}^c$.
Here, we show that a second transition to irreversibility occurs below an $\omega$-dependent critical amplitude, 
$\gamma_{0,\omega}^c \leq \gamma_{0,\phi}^c$, 
in the presence of weak attractions.

\end{abstract}

\maketitle

The flow properties of suspensions remain challenging to predict despite the tremendous progress to date. 
Even the simplest suspensions, consisting of non-Brownian particles suspended in a density-matching Newtonian fluid, 
while exhibiting a Newtonian behavior in steady shear (SS) flow, can show very rich phenomena under oscillatory shear (OS), 
such as flow irreversibility and chaos \cite{Pine_Nature_2005}, 
absorbing state transitions \cite{Corte_NatPhys_2008, Corte_etal_PRL2009}, 
and microstructure reorganizations at large accumulated strains \cite{Bricker_Butler2006, Bricker_Butler2007}. 
All these fundamental behaviors are predicted using very few ingredients in the equations of motion that are in order: 
hydrodynamic forces, including lubrication between adjacent particles, and hard-sphere collisions. 
Because the latter do not possess any characteristic time scale and their amplitude is proportional to the driving hydrodynamic force, 
the system and all its material functions are rate-independent \cite{hinch_2011, chacko_mari_fielding_cates_2018}. 

So far, it has been assumed that the suspension microstructure depends on the strain amplitude ($\gamma_0$) and strain history ($\gamma_{tot}$) in OS,
and rate dependence (if any) is manifested in both SS and OS.
However, recent experiments challenge this assumption and demonstrate a frequency ($\omega$)-dependent rheology in OS 
in the absence of any rate-dependence in SS \cite{Carotenuto_conf_2014,Martone_conf_2018,Martone_experiment}. 
These authors demonstrate that, in OS, the suspension viscosity only depends on the maximum shear rate ($\gamma_0\omega$)
and that data taken at different volume fractions ($\phi$) can be rescaled on a single master curve, 
so to highlight a universal behavior of these materials. 
Furthermore, this rheological observation questions the physics of self-organization in the simplest driven noncolloidal suspension:
it was assumed that suspensions undergo transitions from reversible absorbing states to irreversible chaotic states 
if $\gamma_0$ exceeds a $\phi$-dependent critical amplitude, $\gamma_0^c(\phi)$,
independent of the driving frequency
\cite{Pine_Nature_2005, Corte_NatPhys_2008, Pham_etal_2016}; 
now the experiments in \cite{Martone_experiment} may imply that $\omega$ also affects irreversibility.

In this Letter, we combine experiments with simulations and show that a weak interparticle attraction (\eg van der Waals force) 
is enough to induce the sought rate dependence in OS, while keeping the SS behavior rate-independent. 
Moreover, we reveal that this rheological behavior in OS
 is accompanied by enhanced particle diffusivities and cluster formations 
below a critical shear rate, 
thus the onset of rate dependence when reducing $\gamma_0$ is closely related to the threshold for irreversibility.
A deeper understanding  of self-organization and dynamical phase transition in suspensions is not only of fundamental interest 
\cite{Menon_Ramaswamy2009, echo2009, Royer49, Ness_Cates_PRL2020}, 
but has also attracted practical attention due to its applications in hyperuniform photonic materials or suspension flow control
\cite{Hexner2015, Wilken2020, Nesseaar3296}. 
This letter shows that a physical ingredient up to now neglected, \ie interparticle attraction, must be taken into account.

Our experimental samples consist of glass hollow microspheres (5$\sim$50 $\mu$m in diameter, mean 15.4 $\mu$m) 
dispersed in a Newtonian fluid (polyisobutene, viscosity 15.8 Pa$\cdot$s at 21.2\textdegree{}C) at three volume fractions (20$\sim$40\%).
Time sweep oscillatory tests are executed on a constant-strain rheometer, 
ARES G2 (TA Instruments) equipped with a cone-and-plate geometry, 
by imposing a sinusoidal strain during each run, $\gamma(t)=\gamma_0 \sin(\omega t)$, 
for $\gamma_0$ from 0.5\% to 2 and $\omega$ from 5 to 200 rad/s. 
After a steady preconditioning shear, the complex viscosity is followed in time in units of
the total accumulated strain, 
$\gamma_{tot} = 4 \gamma_0 n_{cyc}$, where $n_{cyc}$ is the number of cycles of the oscillatory shear.
All tests are performed repeatedly according to standard protocols, cf.~Ref.\ \cite{Martone_experiment}.

\begin{figure*}[t]
 \begin{center}
 \includegraphics[width=0.9\textwidth]{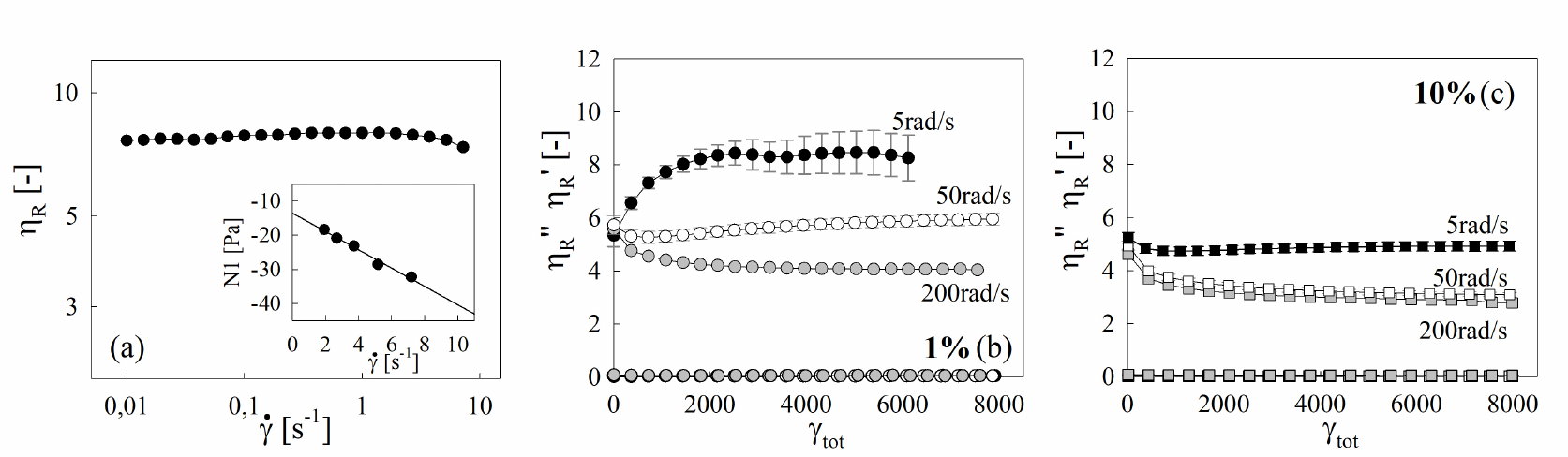}
 \end{center}
 \caption{Experiment at $\phi=40\%$: 
   (a) Relative viscosity ($\eta_R$) and first normal stress difference ($N_1$, inset) vs.\ shear rate in SS.
   (b,c) Evolution of the dynamic viscosity ($\eta'_R$) and its elastic counterpart ($\eta''_R$) vs.\ the total accumulated strain at 
   $\gamma_0=1\%$ (b) and 10\% (c) in OS. The angular frequencies are indicated in the figure.}
 \label{fig:exp raw}
\end{figure*}

The suspensions are interialess and non-Brownian, 
as the particle Reynolds number is smaller than $10^{-6}$ and the P\'eclet number is larger than $10^{5}$. 
Particle sedimentation can be neglected, as the average Shields number is about $10^3$. 
A characteristic time arising from Brownian diffusion or sedimentation is thus irrelevant in the investigated suspensions 
and, accordingly, the SS behaviour shows constant viscosity 
and first normal stress difference negative and linear in the shear rate; see Fig.\ \ref{fig:exp raw}(a). 
In OS, the relative complex viscosity, $\eta_R^* \equiv \eta_R'-i\eta_R''$, 
evolves in time and is a function of $\gamma_0$, 
as widely reported in the literature \cite{Bricker_Butler2006, Breedveld_etal_2001, Corte_NatPhys_2008};
but, surprisingly, it also decreases with $\omega$ (\ie shear thinning).
To illustrate this, we display in Fig.\ \ref{fig:exp raw}(b,c) the relative dynamic ($\eta'_R$) and elastic ($\eta''_R$) viscosities vs.\ $\gamma_{tot}$, 
for two values of $\gamma_0$ and three of $\omega$ at $\phi=40\%$
(data for other $\phi$ are available in Ref.~\cite{Martone_experiment}). 
$\eta'_R$ is always about two orders-of-magnitude larger than $\eta''_R$, 
thus it is practically coincident with $\eta^{*}_R$, highlighting the viscous behavior of the suspension. 
In general, an $\omega$-dependent regime is observed for $\gamma_0$ $\lessapprox 1$ and 
an $\omega$-independent regime otherwise. 
These two regimes coincide with those observed by Lin \etal \cite{Lin_Phan-Thien_Khoo_2013}, 
who showed that in the first regime the microstructure self-arrangement is driven by shear-induced particle diffusions, 
while in the second one the microstructure is immediately formed by the oscillation itself, 
similarly to what happens for a steady flow reversal that is indeed rate-independent. 
The dependence on the frequency and the importance of diffusion indicate that a \emph{non-hydrodynamic} force is at play \cite{drazer_koplik_khusid_acrivos_2002}, 
which cannot be hard-sphere interactions. 
To identify this force, we now turn to numerical simulations.

Discrete element simulations based on a minimal hydrodynamic model are performed for suspensions at zero Reynolds number \cite{hlgd}.
Specifically, our model determines particle nonaffine trajectories according to lubrication, contact (hard-sphere) and interparticle potentials, 
similarly to the Stokesian dynamics \cite{sd1988}.
Since the full many-body hydrodynamic interactions are truncated at the level of lubrication, 
our method is mostly accurate for dense suspensions where interparticle gaps are small, cf.\ Refs.~\cite{Mari_2014JOR, Cheal_Ness_2018}. 
The results below are therefore for $\phi=40\%$ and 50\%
\footnote{All simulations consider bidisperse suspensions of 500 spheres, with size ratio 1.4, in a cubic box subject to Lees-Edwards boundary condition.
In OS, only small amplitudes are used, $\gamma_0 = 0.05 \sim 0.2$, while frequency is controlled by adjusting the non-dimensional time scale.}.

To begin with, we note that several forces have been linked in the literature to rate dependence (shear thinning in particular) in dense, non-Brownian suspensions:
(normal-)load-dependent friction \cite{lobry_lemaire_blanc_gallier_peters_2019},
electrostatic repulsion \cite{Mari_2014JOR, shear_thinning_SM18}
and van der Waals attraction \cite{Brown_nmat2010,Singh_attr_prl2019}.
Friction, as well as any other possible dissipation mechanisms due to surface roughness,
should lead to no less rheological response in SS than OS, 
since larger deformations activate more frictional contacts at the same $\phi$.
Indeed, we have checked that including the contact model of \cite{lobry_lemaire_blanc_gallier_peters_2019} in our system gives shear-thinning in SS instead of OS, 
contrary to what the experiments show.
Therefore, an explanation in terms of friction is unlikely.

As for electrostatic repulsion, its impact on the suspension rheology can be understood as an effective volume effect \cite{shear_thinning_SM18}.
At low shear (thus stress), few particle pairs come closer than an enlarged radius due to the finite-range repulsive force;
the portion of such particles (thus the effective $\phi$) reduces with the shear, leading to shear thinning.
Although plausible, it is difficult to rigorously establish this argument in all scenarios.
In fact, our simulations show that electrostatic repulsion causes shear thickening in OS, 
again contradicting the experimental results \cite{submission}.

The only possibility left is attraction.
Here, we consider the simplest van der Waals (vdW) attraction, 
which is always present regardless of particle size \cite{Israelachvili_book,mewis_wagner_book}.
As in Ref.~\cite{Singh_attr_prl2019}, we model a nonretarded and additive vdW as
$F_{vdw}= A \bar{a}/12(h^2+\epsilon^2)$,
where $A$ is the Hamaker constant, 
$\bar{a}=2a_ia_j/(a_i+a_j)$ the harmonic mean radius of two interacting particles,
$h$ their surface gap and $\epsilon$ a small constant to prevent divergence of $F_{vdw}$ at $h=0$;
we use $\epsilon=\sqrt{10^{-5}}a$ to model glass beads.
The maximal attraction $\mathcal{F} =Aa/12\epsilon^2$ introduces a characteristic stress scale, $\uptau_\mathcal{F} = \mathcal{F}/\pi a^2$,
for two touching particles,  opposed to the minimal shear stress,  $\uptau_s=\eta_0\gamma_0\omega/2\pi$ ($\eta_0\dot{\gamma}$ in SS),
where $\eta_0$ denotes the solvent viscosity.
The ratio $\uptau_s/\uptau_\mathcal{F}$ thus defines a non-dimensional shear rate, hereafter denoted
Sr $\equiv 6\eta_0\epsilon^2 a \gamma_0\omega/A$.

\begin{figure}[t]
 \begin{center}
 \includegraphics[width=0.8\columnwidth]{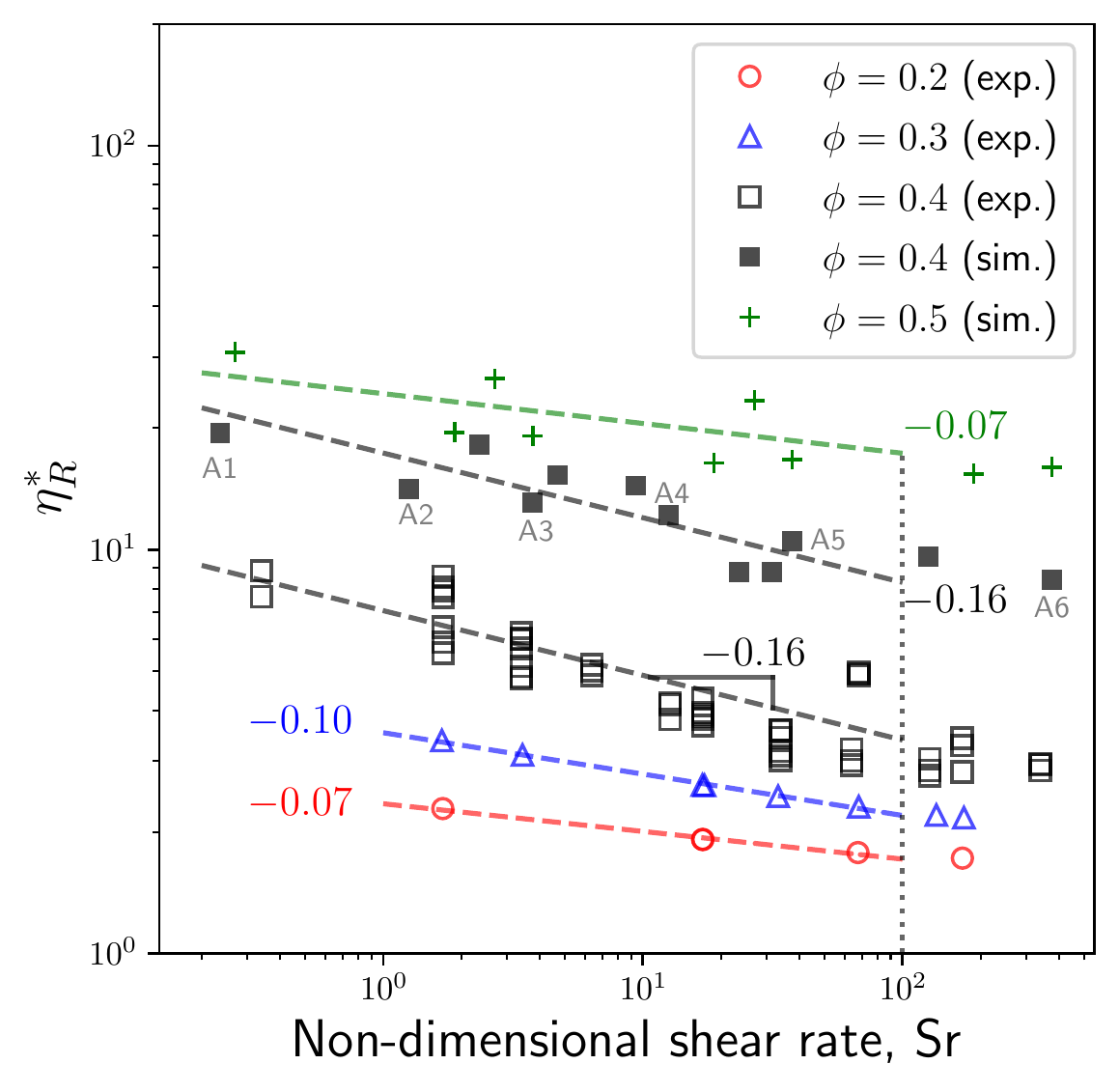}
 \begin{picture}(0,0)
      \setlength{\unitlength}{\columnwidth}
      \put(-0.68,0.54){\includegraphics[height=1.8cm]{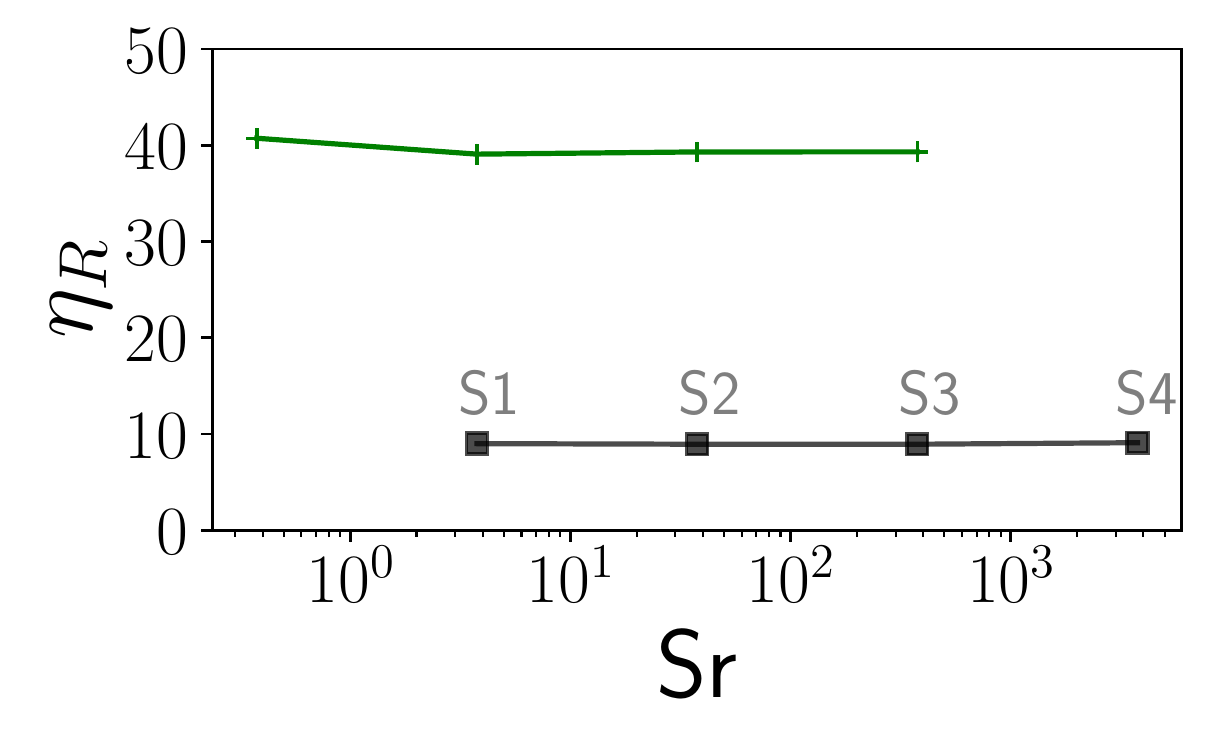}}
    \end{picture}
 \end{center}
 \caption{Suspension viscosities in OS and SS (inset). 
 Dashed lines are least squares fits to data for Sr $<100$; numbers indicate the slopes.}
 \label{fig:visc}
\end{figure}

Fig.\ \ref{fig:visc} shows the rheology of this model adhesive system.
Remarkably, with the addition of only a weak vdW attraction ($\propto$$1/$Sr), 
the essential features of the complex rate-(in)dependent behaviors in (SS)OS seen in experiments are captured numerically
\footnote{The vertical discrepancy between simulations and experiments may be due to polydispersity of the glass beads 
or residual short-range repulsion forces that are not modeled numerically.}.
Specifically, below a critical shear rate in OS, $\eta_R^*$ exhibits power-law reductions with shear, 
$\eta_R^* \sim$ Sr$^{-\alpha}$, 
with $\alpha$ a $\phi$-dependent positive exponent largest at $\phi=40\%$.  
This non-monotonic dependence can be understood by considering the limits of very dilute and highly-packed suspensions: 
at vanishing volume fractions, collisions and vdW attractions are negligible, 
the system is dominated by hydrodynamic forces and therefore nearly rate-independent; 
close to packing, lubrication and contact forces dominate but, 
because both are proportional to the shear rate,
the system again exhibits weak rate dependence.

\begin{figure}[t]
 \begin{center}
 \includegraphics[width=.9\columnwidth]{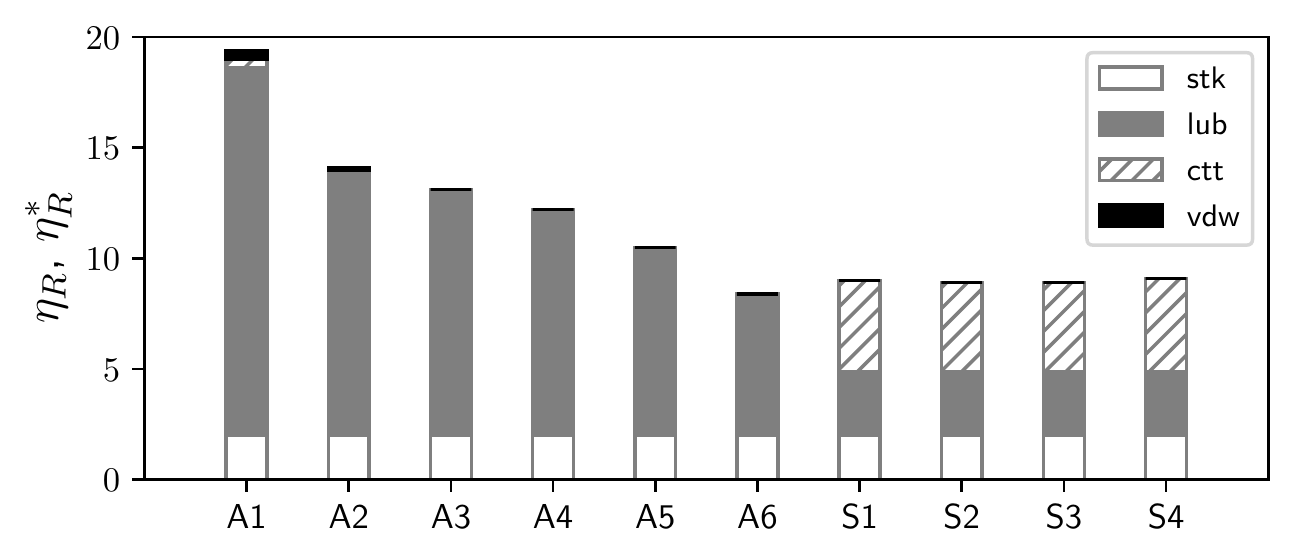}
 \end{center}
 \caption{Shear stress components of representative OS and SS cases at $\phi=40\%$; 
 see Fig.\ \ref{fig:visc} for the annotation.}
 \label{fig:budget}
\end{figure}

To provide quantitative evidence, we display the different contributions to the relative viscosity for ten representative cases in Fig.~\ref{fig:budget}.
The budget terms correspond to the Stokes (stk), lubrication (lub), contact (ctt) and vdW forces extracted from the simulations \cite{hlgd}.
Surprisingly, at $\phi=40\%$, where the maximum vdW effect is expected,
virtually no contributions from the attractive vdW forces are visible across three decades of Sr. 
In SS, the lubrication and contact stresses are constant
(stk only depends on $\phi$ thus identical in all cases) and so the relative viscosity; 
in OS, however, lubrication is shear-dependent while contact stresses are vanishing.
Since the rate dependence results from vdW,
naively, one would expect it also makes a rate-dependent contribution to the stress budget, 
or at least be active in OS. 
Its absence implies that the rheology is \emph{indirectly} modulated by weak attractions.
The question, then, is how such indirect modulation occurs.

Recall that, without any inertial, thermal or non-Newtonian effects, 
the rheological properties of a suspension at any given time are determined solely from its underlying microstructure \cite{mewis_wagner_book}.
The microstructure evolution is 
mediated by time-reversible Stokes flows,
though the particle dynamics themselves satisfy Onsager's variational principle for general irreversible processes \cite{Doi2011}.
Indeed, both experiments and simulations have shown that particle diffusion can occur in slowly sheared suspensions,
leading to irreversible and ultimately chaotic dynamics 
\cite{Eckstein1977, leighton_acrivos_1987, Breedveld2001, drazer_koplik_khusid_acrivos_2002, Pine_Nature_2005}.
Particularly, in OS, the threshold for irreversibility was found to be 
gauged by a $\phi$-dependent critical strain amplitude $\gamma_0^c(\phi)$:
for $\gamma_0 > \gamma_0^c$, suspensions are irreversible with non-vanishing effective diffusivities, 
$D_i = \langle (\Delta_i/a)^2 \rangle /2\gamma_{tot}$ 
($\Delta_i$ denotes the particle displacement in any spatial direction $i$ at integer-period intervals); 
for $\gamma_0 < \gamma_0^c$, the dynamics evolve towards absorbing states with $D_i \approx 0$ in finite time.
The system undergoes a continuous phase transition 
(likely conserved directed percolation \cite{Ness_Cates_PRL2020}) 
at $\gamma_0 = \gamma_0^c$, 
indicated by a nonzero order parameter $\langle f_a^\infty \rangle$ based on the mean fraction of actively colliding particles.
This theoretical framework \cite{Corte_NatPhys_2008} applies to a remarkable variety of situations, 
see \eg Ref.~\cite{Ness_Cates_PRL2020} and references therein.
We note that none of them invokes a role for frequency, though.

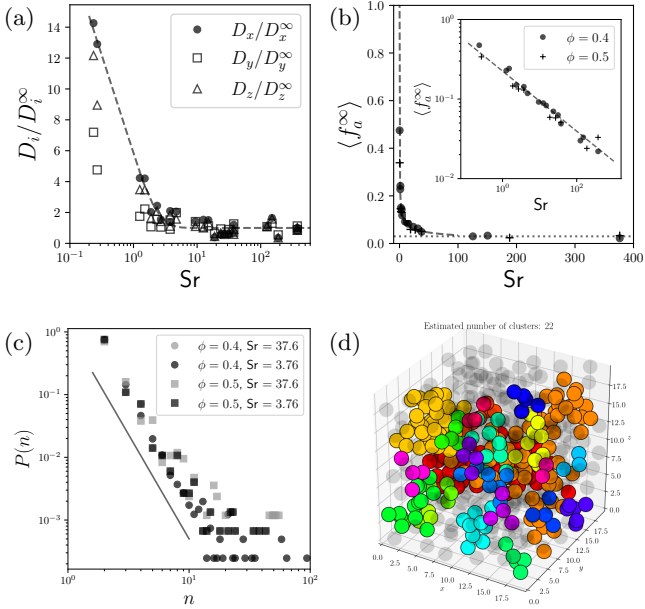
\begin{figure}[t]
 \begin{center}
 \includegraphics[width=0.48\columnwidth]{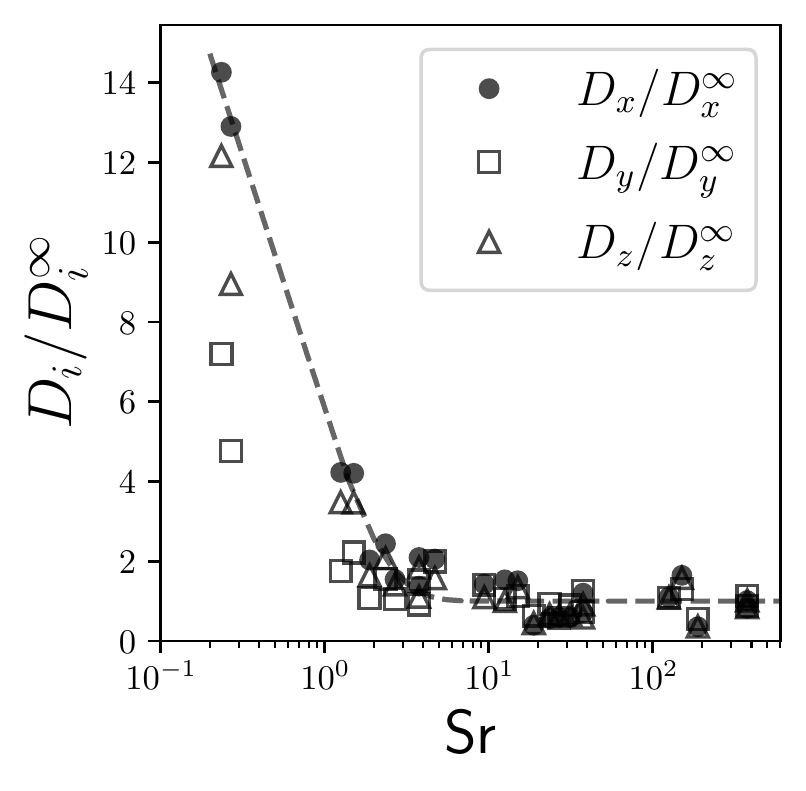}
 \includegraphics[width=0.50\columnwidth]{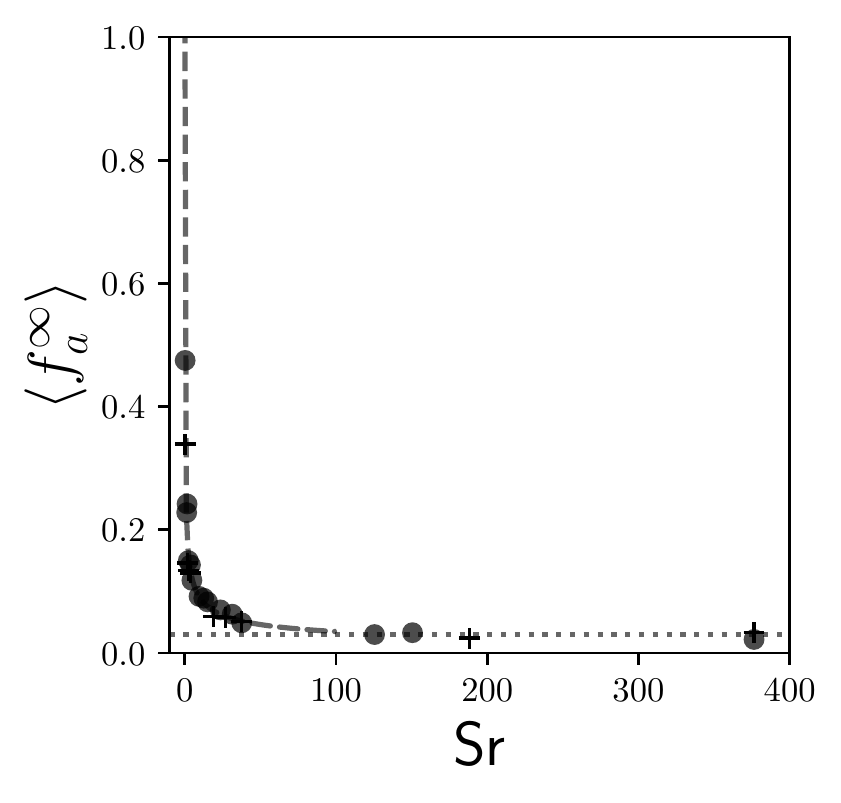}
 \includegraphics[width=0.49\columnwidth]{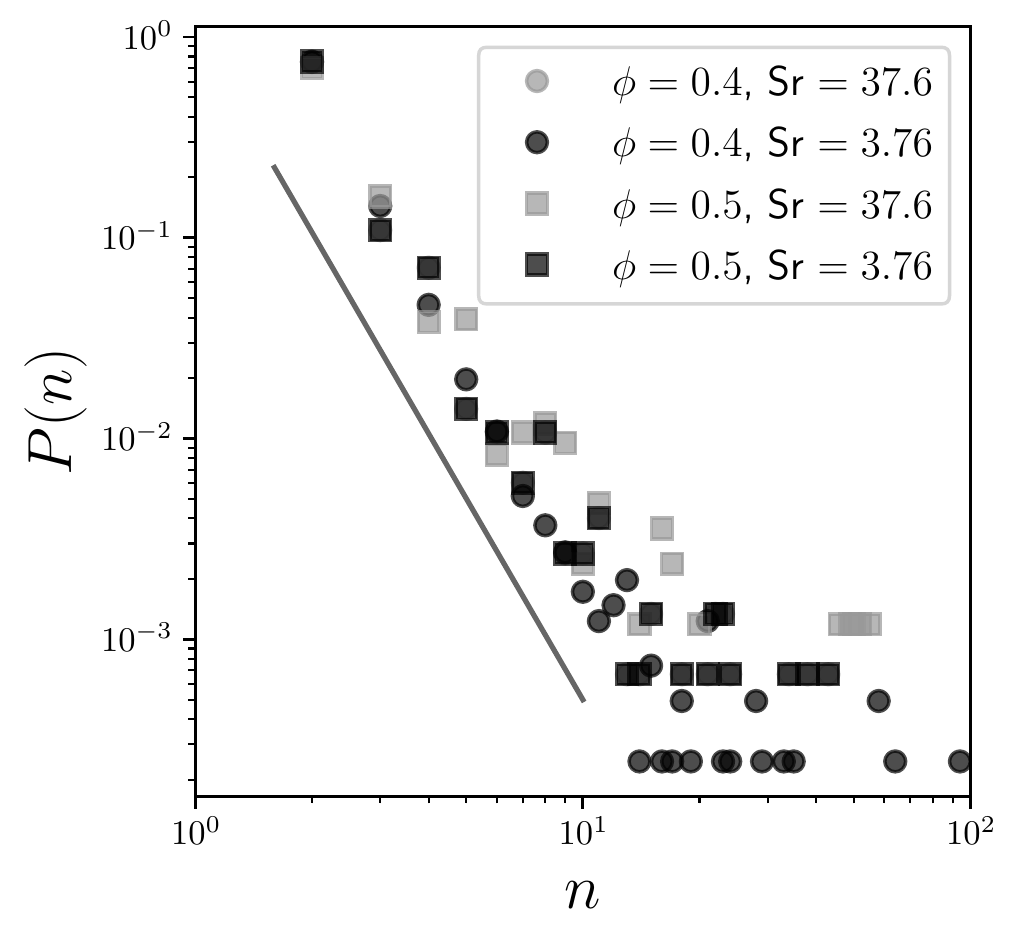}
 \includegraphics[width=0.48\columnwidth]{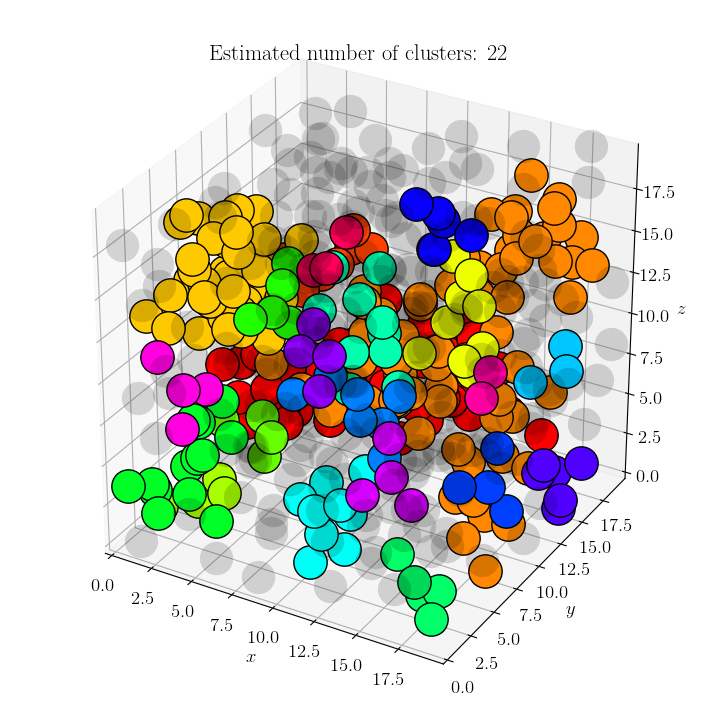}
  \begin{picture}(0,0)(123,10)
      \setlength{\unitlength}{\columnwidth}
      \put(0.12,0.65){\includegraphics[height=2.8cm]{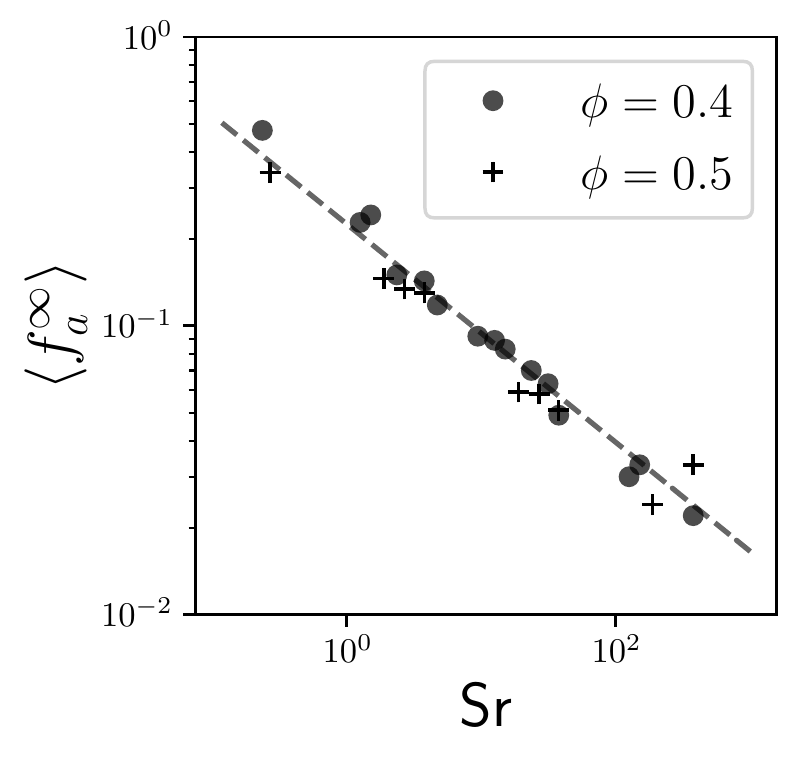}}
      \put(-0.5,0.95){(a)} \put(0,0.95){(b)} \put(-0.5,0.45){(c)} \put(0,0.45){(d)}
  \end{picture}
 \end{center}
 \caption{Microstructure statistics in OS.
 (a) Relative particle diffusivities at $\phi=$ 40\% and 50\% (not distinguished). The dashed line is a fit of $D_x/D_x^\infty$.
 (b) Mean fraction of active (colliding) particles on linear and logarithmic (inset) scales. The dashed line is a power-law fit with slope $\beta=-0.40$.
 (c) Power-law distribution of cluster size, $P(n) \sim n^{-3.33}$ (solid line).
 (d) A snapshot of the particle suspension at $\phi=$ 40\%. 
 Different cluster groups are indicated by colors.}
 \label{fig:stat}
\end{figure}

It is therefore natural to examine the microstructure statistics of our suspensions.
Fig.~\ref{fig:stat} shows the numerical results at $\phi=40$\% and 50\%.
Surprisingly, at strain amplitudes well under the irreversibility threshold just mentioned
($\gamma_0 \leq 0.2$ for all simulations with  $\gamma_0^c \approx 0.82$ at $\phi=$40\% 
\footnote{$\gamma_0^c(\phi)$ can be calculated from the power-law fit $C\phi^{-\alpha}$ reported in Ref.~\cite{Pine_Nature_2005}.
However, there is a typo for the value of $\alpha$. It should be 1.93 instead of $-1.93$.}),
our suspensions can be both irreversible and active.
This irreversibility only occurs below a fixed critical shear rate, Sr$_c$.
Fig.~\ref{fig:stat}(a) shows that, 
by renormalizing the effective diffusivities with their averages in the infinite shear limit, 
\eg $D_x^\infty = \langle D_x \rangle |_{\textrm{Sr}\to \infty}$,
all values at $\phi=$ 40\% and 50\% diverge with reducing Sr.
A similar behavior is observed for the order parameter, 
which diverges below Sr$_c$ as $\langle f_a^\infty \rangle \sim$ Sr$^\beta$, 
with $\beta=-0.40$ (Fig.~\ref{fig:stat}b).
Akin to the self-organized criticality due to slow sedimentation \cite{Corte_etal_PRL2009}, 
particles in adhesive suspensions actively form clusters with a near power-law size distribution (Fig.~\ref{fig:stat}c).
As an example, Fig.~\ref{fig:stat}(d) shows a snapshot of the suspension at $\phi=40$\%, 
where various particle clusters are displayed in color
(inactive particles are displayed in grey).

Comparison of Figs.~\ref{fig:stat}(a,b) and \ref{fig:visc} indicates that 
this new threshold for irreversibility is closely related to the onset of rate dependence,
both at Sr$_c \approx 100$ 
\footnote{In general, the onset of rate dependence may depend on $\phi$ through a critical suspension shear stress,
$\sigma^* \sim \eta_R^*(\phi)$Sr$_c$, 
since rheological quantities average over the entire suspension.
In our case, $\eta_R^*(\phi)$ differs at most by a factor of 4 at large Sr, 
thus we refrain from introducing a $\phi$ dependence in Sr$_c$ for simplicity.}.
Below Sr$_c$, the microstructure is indirectly modulated by weak vdW interactions; 
their 
relative intensity is inversely proportional to the shear rate,
thus the suspension shows diffusive dynamics and rate dependence at low $\gamma_0$.
Above Sr$_c$, hydrodynamic interactions overcome vdW, 
resulting in time-reversible Stokesian dynamics and a rate-independent suspension rheology.
Here, we cannot determine the value of Sr$_c$ exactly 
due to the increasing uncertainties in the data near it.
Nevertheless, by matching experiments and simulations at Sr$_c=100$ (Fig.~\ref{fig:visc}),
we estimate the Hamaker constant of the vdW interaction to be $A \approx 4\times 10^{-19}$J in the system under investigation. 
This is within the range of $A$ for most condensed phases (cf.~Ref.~\cite{Israelachvili_book}, page 254),
which validates our assumption.

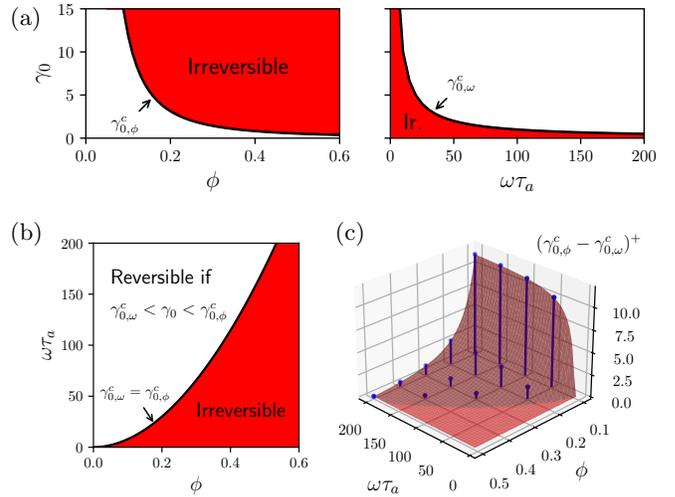
\begin{figure}[t]
 \begin{center}
 \includegraphics[width=\columnwidth]{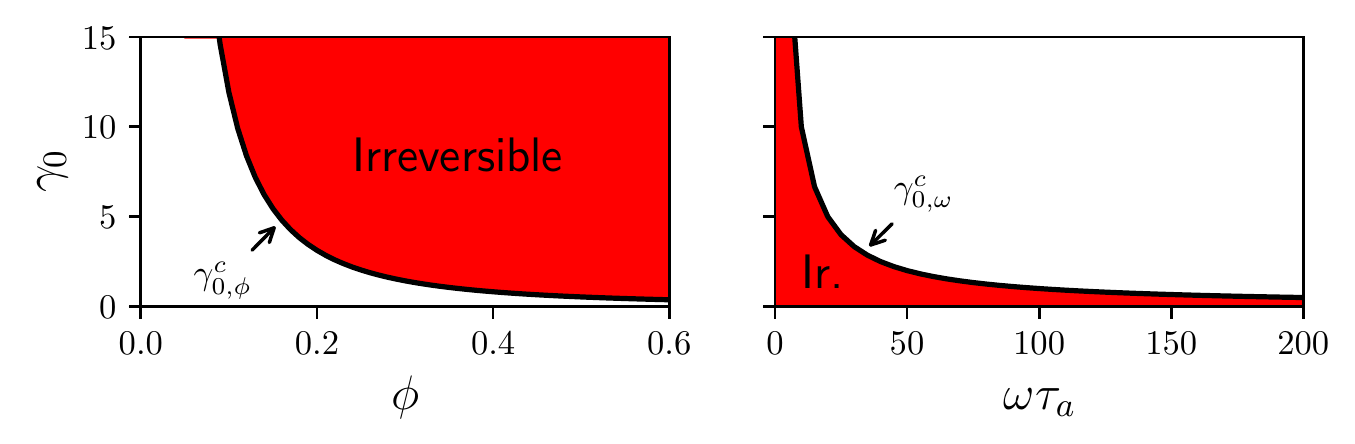}
 \includegraphics[width=0.45\columnwidth]{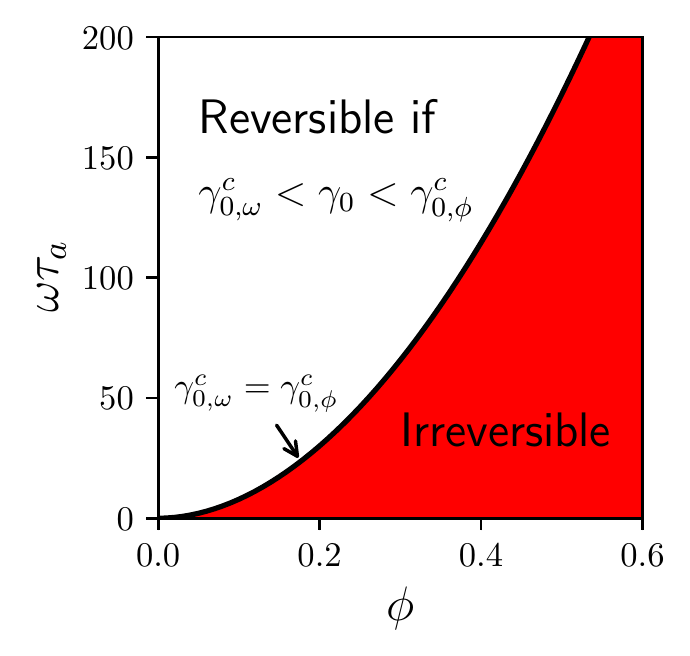}
 \includegraphics[width=0.5\columnwidth]{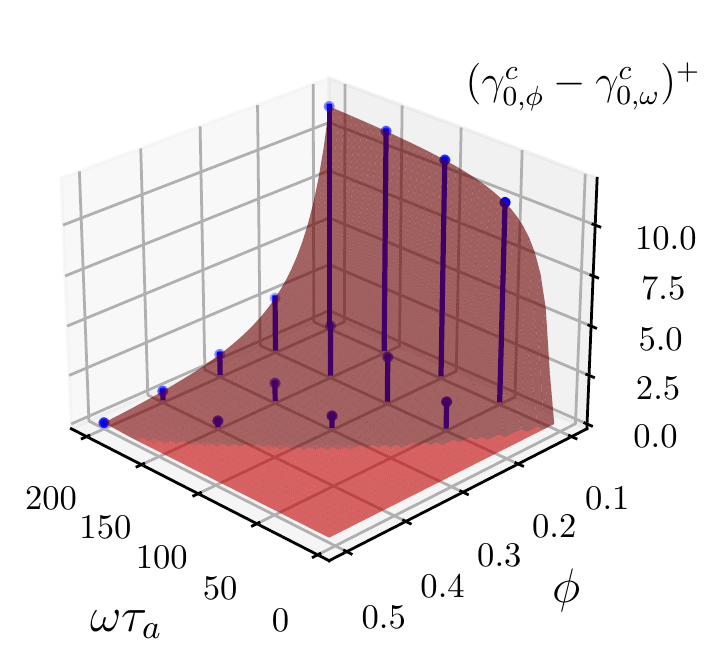}
  \begin{picture}(0,0)(123,10)
      \setlength{\unitlength}{\columnwidth}
      \put(-0.5,0.78){(a)} \put(0,0.45){(c)} \put(-0.5,0.45){(b)}
  \end{picture}
 \end{center}
 \caption{Phase diagrams for irreversibility in sheared adhesive suspensions.
 (a) Critical strain amplitudes due to collisions (left) or attractions (right).
 $\omega$ is normalized by $\tau_a \equiv 6\eta_0\epsilon^2 a/A$.
 (b) Projection of the irreversibility map on the $(\phi,\omega\tau_a)$ plane.
 (c) Volume of the reversible absorbing states in the $(\phi,\omega\tau_a,\gamma_0)$ space.}
 \label{fig:phase}
\end{figure}

More importantly, the existence of a unique Sr$_c$ suggests that 
there is a \emph{frequency}-dependent critical amplitude for the transitions between absorbing and diffusing states.
This can be written as $\gamma_{0,\omega}^c =$ Sr$_cA/6\eta_0\epsilon^2 a \omega$,
which is independent of the volume fraction and thus must coexist with $\gamma_0^c(\phi)$
(hereafter $\gamma_{0,\phi}^c$).
In between the two thresholds,
\ie $\gamma_{0,\omega}^c < \gamma_0 < \gamma_{0,\phi}^c$,
suspensions can reach reversible absorbing states from any initial condition; 
below and above this range,
irreversibility arises from either attractions (low $\gamma_0$) or collisions (high $\gamma_0$); 
see Fig.~\ref{fig:phase}(a,b).
However, $\gamma_{0,\omega}^c$ may also exceed $\gamma_{0,\phi}^c$, 
resulting in irreversible dynamics for all $\gamma_0$.
For example, consider a 40\% suspension oscillating at 5 rad/s:
if $\gamma_0 > 0.82$, the suspension is chaotic due to collisions \cite{Pine_Nature_2005, Corte_NatPhys_2008};
whereas if $\gamma_0 \leq 0.82$, it is irreversible due to attractions 
($\gamma_{0,\omega}^c=18.8$ for our suspensions).
Note that, when $0.82 < \gamma_0 < 18.8$, both these mechanisms promote irreversibility.
In such a case, the dynamics are controlled by collisions 
because collision-induced diffusions at larger $\gamma_0$ occur over a shorter time scale.
We have checked that the suspension relaxation time (\ie the time to reach steady states) is always orders-of-magnitude longer
in our simulations and experiments (cf.~Fig.~\ref{fig:exp raw}) than those in Ref.~\cite{Corte_NatPhys_2008}.
Thus, for large values of $\gamma_0$ the system dynamics are collision-dominated: 
this finally explains why rate dependence is observed only for smaller $\gamma_0$ in OS, but not in SS, 
under the same shear rate.

A new phase diagram thus emerges for irreversibility in sheared adhesive suspensions.
Fig.~\ref{fig:phase}(b,c) shows the region of reversible steady dynamics in $\phi$ and $\omega$, 
and the size of the reversibility window in $\gamma_0$;
the latter is measured by a positive difference between the two critical amplitudes, 
$(\gamma_{0,\phi}^c - \gamma_{0,\omega}^c)^+$,
where $(x)^+=\max(x,0)$.
Notice how rapidly the window reduces as $\phi$ or $1/\omega$ increases.
Considering the chaotic nature of many-body systems (\eg ideal gas, planets, etc.), 
microscopic irreversibility is rather the hallmark than an exception for suspensions.  
This is true even in Stokes flow for, whenever particles get close, 
their dynamics are not governed by the hydrodynamics alone. 
Here, by combining experiments with simulations, 
we have shown that there is a fundamental connection between irreversibility and suspension rheology, 
both of which depend on the strain amplitude as well as the driving frequency. 
Future work may examine this picture when including long-range hydrodynamic interactions,
or explore the so-called active fluids whose intriguing rheologies defy reversibility even without passive particles \cite{Saintillan2018}.

We thank R.~Radhakrishnan, C.~Ness, F.~Peters, A.~Los, J.~Chun and A.~Leshansky for helpful discussions.
The work is supported by the Swedish Research Council (grant no.\ VR 2014--5001)
and University of Campania `L. Vanvitelli' under the  programme ``VALERE: VAnviteLli pEr la RicErca'' project: SEND.

\bibliographystyle{apsrev4-2}
\small
\bibliography{main}



\end{document}